\documentclass[12pt,preprint]{aastex}
\usepackage{amsmath}

\shorttitle{Active asteroid P/2016 J1}
\shortauthors{Moreno et al.}

\begin{document}


\title{The splitting of double-component active asteroid P/2016 J1 (PANSTARRS)}


\author{F. Moreno\affil{Instituto de Astrof\'\i sica de Andaluc\'\i a, CSIC,
  Glorieta de la Astronom\'\i a s/n, 18008 Granada, Spain}
\email{fernando@iaa.es}}
\author{F.J. Pozuelos\affil{Instituto de Astrof\'\i sica de Andaluc\'\i a, CSIC,
  Glorieta de la Astronom\'\i a s/n, 18008 Granada, Spain} }
\author{
B. Novakovi\'c\affil{Department of Astronomy, Faculty of Mathematics,
  University of Belgrade, 
Studentski trg 16, 11000 Belgrade, Serbia}}
\author{
J. Licandro\affil{Instituto de Astrof\'\i sica de Canarias,
  c/V\'{\i}a 
L\'actea s/n, 38200 La Laguna, Tenerife, Spain, 
\and 
 Departamento de Astrof\'{\i}sica, Universidad de
  La Laguna (ULL), E-38205 La Laguna, Tenerife, Spain}}  
\author{
A. Cabrera-Lavers\affil{GRANTECAN, Cuesta de San Jos\'e s/n, E-38712 ,
  Bre\~na Baja, La Palma, Spain, 
\and
Instituto de Astrof\'\i sica de Canarias,
  c/V\'{\i}a 
L\'actea s/n, 38200 La Laguna, Tenerife, Spain}}
\author{
Bryce Bolin\affil{
Laboratoire Lagrange, Universit\'e C\^ote d'Azur, Observatoire de la
C\^ote d'Azur, CNRS, Blvd. de l'Observatoire, CS 34229, 06304 Nice
cedex 4, France}}
\author{
Robert Jedicke\affil{University of Hawaii, Institute for Astronomy, 2680 
Woodlawn Dr, Honolulu, HI, 96822}}
\author{
Brett J. Gladman\affil{Department of Physics and Astronomy, University
  of British Columbia, 
Vancouver, BC, Canada}}
\author{
Michele T. Bannister\affil{Astrophysics Research Centre, Queens
  University Belfast, BT7 1NN, 
Northern Ireland, UK}}
\author{
Stephen D. J. Gwyn\affil{NRC-Herzberg Astronomy and Astrophysics,
  National Research Council of 
Canada, 5071 West Saanich Rd, Victoria, British Columbia V9E 2E7,
Canada}}
\author{
Peter Vere\v{s}\affil{Jet Propulsion Laboratory, California Institute of Technology, 4800 Oak Grove Drive, Pasadena, CA 91109, USA}}
\author{
Kenneth Chambers\affil{Institute for Astronomy University of Hawaii,
  2680 Woodlawn Dr, 
Honolulu, HI, 96822, USA}}
\author{
Serge Chastel\affil{Institute for Astronomy University of Hawaii,
  2680 Woodlawn Dr, 
Honolulu, HI, 96822, USA}}
\author{
Larry Denneau\affil{Institute for Astronomy University of Hawaii,
  2680 Woodlawn Dr, 
Honolulu, HI, 96822, USA}}
\author{
Heather Flewelling\affil{Institute for Astronomy University of Hawaii,
  2680 Woodlawn Dr, 
Honolulu, HI, 96822, USA}}
\author{
Mark Huber\affil{Institute for Astronomy University of Hawaii,
  2680 Woodlawn Dr, 
Honolulu, HI, 96822, USA}}
\author{
Eva Schunov\'a-Lilly\affil{Institute for Astronomy University of Hawaii,
  2680 Woodlawn Dr, 
Honolulu, HI, 96822, USA}}
\author{
Eugene Magnier\affil{Institute for Astronomy University of Hawaii,
  2680 Woodlawn Dr, 
Honolulu, HI, 96822, USA}}
\author{
Richard Wainscoat\affil{Institute for Astronomy University of Hawaii,
  2680 Woodlawn Dr, 
Honolulu, HI, 96822, USA}}
\author{
Christopher Waters\affil{Institute for Astronomy University of Hawaii,
  2680 Woodlawn Dr, 
Honolulu, HI, 96822, USA}}
\author{
Robert Weryk\affil{Institute for Astronomy University of Hawaii,
  2680 Woodlawn Dr, 
Honolulu, HI, 96822, USA}}
\author{
Davide Farnocchia\affil{
Jet Propulsion Laboratory, California Institute of Technology, 4800 Oak Grove Drive, Pasadena, CA 91109, USA}}

\and

\author{
Marco Micheli\affil{
SSA-NEO Coordination Centre, ESA, 00044 Frascati (RM), Italy,
\and
INAF OAR, 00040 Monte Porzio Catone (RM), Italy
}}


\begin{abstract}

We present deep imaging observations, orbital dynamics, and dust tail
model analyses of the double-component asteroid P/2016 J1 (J1-A and
J1-B). The observations were acquired at 
the Gran Telescopio Canarias (GTC) and the 
Canada-France-Hawaii Telescope (CFHT) from mid March to late July, 2016. 
A statistical analysis of backward-in-time integrations of the orbits 
of a large sample of clone objects of P/2016 J1-A and J1-B   
shows that the minimum separation between them 
occurred most likely $\sim$2300 days prior to
the current perihelion passage, i.e., during the previous orbit 
near perihelion.  This closest approach was probably 
linked to a fragmentation event of their parent body. Monte Carlo dust
tail models show that  
those two components became active simultaneously $\sim$250 
days before the current perihelion, with comparable maximum loss rates of $\sim$0.7 kg
s$^{-1}$ and $\sim$0.5 kg s$^{-1}$, and total ejected masses of
8$\times$10$^{6}$ kg  and  6$\times$10$^{6}$ kg for fragments
J1-A and J1-B, respectively. In consequence, the fragmentation event and the
present dust activity are unrelated. The simultaneous activation times of
the two components and the fact that the activity lasted 6 to 9 months
or longer, strongly indicate ice sublimation as the most likely 
mechanism involved in the dust emission process.   

\end{abstract}


\section{Introduction}
The double-component asteroid P/2016 J1 (PANSTARRS) (components 
designated as J1-A and J1-B) was discovered by R. Weryk and R. J. Wainscoat 
on CCD images acquired on May 5.5 UT with the 
1.8-m Pan-STARRS1 telescope \citep{Weryk16}. The object is classified
as a main belt asteroid because its Tisserand parameter with respect
to Jupiter  \citep{Kresak82} is $T_J$=3.113 (most main belt asteroids
have $T_J >$3). Up
to date, some 
twenty objects in typical asteroidal orbits have been found 
showing transient comet-like 
appearance. The first object 
of this kind, 133P/Elst-Pizarro,  
was discovered in 1996, and since then it has shown alternate periods
of activity and inactivity 
\citep[e.g.,][]{Hsieh04,Hsieh10,Jewitt14a}. The orbits of these
objects are found to be generally stable over timescales
longer than those of Jupiter-family comets or Halley-type comets, so
that they are very likely native to the asteroid belt, and not interlopers from
the outer solar system \citep[see
  e.g.,][]{Haghighipour09,Hsieh13}. 

A variety of activation
mechanisms for these objects have been 
proposed, from impact-induced to rotational disruption (for most of
the short-duration events), to ice-sublimation (when the activity
lasts typically a few months, in which case they are sometimes named 
main-belt comets). For reviews of the objects
found so far, and their proposed activation mechanisms, see 
\cite{Bertini11,Jewitt15}.

The case of P/2016 J1 is remarkable as it is the first 
time that a double-component active asteroid sharing very similar orbital
elements and patterns of activity has been discovered. In this paper, we
first report a dynamical study of the orbital evolution of the two
components by  
backward in time numerical integration 
of their orbits in order to assess their common origin, and the
fragmentation time of the parent body. And second, we 
characterize the activity pattern of the two components by the
photometric fit to the dust tails during the four months and a half
spanned by the observations.

\section{Observations and data reduction}

Observations of P/2016 J1 were scheduled within our GTC
program of observations after the discovery alert. Images 
of P/2016 J1 have been obtained under photometric conditions on
the nights of 14 May, 28 May, and 31 July, 2016. The
images were  obtained on a CCD using a Sloan $r^\prime$ filter in the
Optical System for Image and Low Resolution Integrated Spectroscopy
(OSIRIS) camera-spectrograph \citep{Cepa00,Cepa10} at the GTC. The
plate scale was 0.254 $\arcsec$/pixel. The images were bias subtracted,
flat-fielded, and calibrated using standard stars. A median stack image
was produced each night of observation from the available
frames (see Figure 1). 

In addition, the object was serendipitously recorded on March 17,
2016, 14:53 UT, on the MegaCam detector \citep{Boulade03} 
of the 3.6m CFHT, found using
Solar System Object Image  
Search (SSOIS) \citep{Gwyn12}, in an image taken as part of the Outer
Solar System Origins Survey (OSSOS) \citep{Bannister16}. Additional
data from the Pan-STARRS telescope taken between 
2016-03-04 and 2016-05-05 were used to refine the orbit of and locate
P/2016 J1 in the CFHT SSOS database. The MegaCam 
detector provides 1-degree wide images on a mosaic of 40 CCDs at a
scale of 0.184 \arcsec/pixel. This image was obtained through 
the wide-band gri.MP9605 filter, using sidereal tracking, 
implying that the asteroid components appear
trailed because of their motion on the sky (see Figure 1). Thus, it was
not possible to retrieve the isophote field, although
their magnitudes were determined to constraint the dust activity
model.

The log of the observations is presented in Table
1. This table includes relevant geometric parameters of the 
observations and the magnitudes of the two
components. Those magnitudes are all computed on apertures of 5000 km radius
projected on the sky, and converted to magnitudes in the 
the standard $R_c$ Cousins/Bessell band. We assume a solar-like
spectrum for the scattered light from 
the asteroid dust. Since the asteroid is located in the outer belt,
where C-type asteroids are abundant, this is consistent with the
featureless, flat 
spectra, shown by those objects at wavelengths longer than 400nm
\citep[see, e.g.][]{Pater10}.  The conversion from  Sloan-$r^\prime$ to $R_c$
magnitudes was made 
by $R_c$=$r^\prime$--0.19, obtained by assuming $(V-R_c)_\odot$=0.354
 \citep{Holmberg06}, and the photometric relation 
$r^\prime = V-0.84(V-R_c)+0.13$ \citep{Fukugita96} with $V=V_\odot$=--26.75
 \citep{Cox00}. The conversion from gri.MP9605 to $R_c$ magnitude was
performed 
through convolution of the solar spectrum \citep{Neckel84} with the
bandpasses of the corresponding filters, resulting in
$R_c$=gri.MP9605+1.02. 

In all cases, both asteroid components appear active at the time of the
observations, so that only upper limits to the nuclear sizes can be
provided. We obtain the absolute magnitudes $H$ from the $V$ magnitudes  
using the \cite{Bowell89} formalism, for which we 
assume a slope parameter of $G$=0.15, appropriate for C-type asteroids of the
outer belt. We then apply the $H$-diameter relationship by
\cite{HarrisLagerros02}. The highest H-magnitude for J1-A fragment is
$H$=19.22$\pm$0.12 (March 17, 2016), whereas for J1-B it is
$H$=19.35$\pm$0.03 (July 31, 
2016). This would translate to maximum diameters of $\sim$1000 and  
$\sim$900 m,
for J1-A and J1-B, respectively, 
assuming a geometric albedo of $p_v$=0.04, appropriate for C-type
asteroids. Assuming  
bulk densities in the 1000 to 3000 kg m$^{-3}$ range this
would imply escape velocities from 0.37 to
0.65 m s$^{-1}$, and from 0.34 to
0.58 m s$^{-1}$ for components J1-A and J1-B, respectively.

\section{Orbital dynamics simulations}

In order to assess the probable common origin of the two asteroid components we 
analyzed their possible past orbital histories. For this task we used
the Orbit9 integrator embedded in the OrbFit
package.\footnote{http://adams.dm.unipi.it/orbfit/} The orbits were  
propagated backward in time for about 100 years (i.e. about 18
revolutions around the Sun),  
starting from the date of the most recent perihelion passage (June
24th 2016). To explore the statistically possible orbital
configurations we generated 
$4\times10^{4}$ clone combinations drawn from the multivariate normal
distribution, which is defined by the orbital covariance matrix. All
the orbital data and their 
uncertainties are taken from the JPL Small Bodies database.  
The dynamical model includes as perturbing bodies all the major
planets, while the clones 
were treated as massless particles. This analysis neglected
non-gravitational perturbations,  
which are likely dominated by the current orbital uncertainties due to
the short observed arcs used in estimating the trajectories of the two objects.

For each pair of clones, we obtained a time evolution of their mutual
distances, and 
recorded the instants of the closest approach. The results obtained
for all clones are shown in  
the upper panel of Figure 2. They suggest two possible solutions for
the age of this pair,  
i.e. it should be either about 900 or 2300 days old (counting from June 24th 2016). 

Still, the targeting minimum distance is related to the radius of a Hill sphere that characterize 
strength of the mutual gravitational interaction. For the two components studied here the Hill 
radius $r_{Hill}$ is only about 300 km. Therefore, to better access a possible separation date 
we focused on approaches within $5r_{Hill}$, i.e. about 1500 km.

The results taking into account only approaches within $5r_{Hill}$ are
shown in lower panel of Figure 2. 
There is one striking difference between the results obtained for all clones and only with those
that had deep close approach. The most recent of two possible age solutions has disappeared, leaving
the one about 2300 days before perihelion as the only viable
option. Therefore, taking into account only 
close encounters observed around 2300 days before perihelion, we found
a refined estimate that separation event 
occurred $2300\pm270$ days prior to June 24th 2016.

To summarize, the obtained results strongly support the common origin
of two components of P/2016 J1, 
suggesting that a separation event likely occurred about 6 years
ago. This implies that the current  
activity is not a direct consequence of the separation event. 

The two components J1-A and J1-B are also a very interesting example of a population of the 
so-called asteroid pairs \citep{VokNes2009,Milanietal2010}, being the
youngest pair known so far.
 
\section{The Monte Carlo Dust Tail Model}

To perform a theoretical interpretation of the activity pattern
associated to the asteroid components, in terms of the dust 
physical parameters, we used our Monte Carlo dust 
tail code. This code has been used previously on several works on
activated asteroids and comets, including comet
67P/Churyumov-Gerasimenko, the Rosetta 
target \citep[e.g.,][]{Moreno16a}. This model computes the dust tail
brightness of a comet or activated asteroid by adding up the
contribution to the brightness of each particle ejected from the
parent nucleus, that, in the presence of the solar radiation pressure and
gravity forces, follows a Keplerian trajectory. For a
description of the code, \cite[see e.g.,][]{Moreno12a,Licandro13,Moreno16a}. The  ratio of radiation
pressure to the gravity force exerted on each particle is given by
the parameter $\beta =C_{pr}Q_{pr}/(2\rho r)$,  where
$C_{pr}$=1.19$\times$ 10$^{-3}$ kg m$^{-2}$, $Q_{pr}$ is the radiation
pressure coefficient, and $\rho$ is the particle density. $Q_{pr}$ 
is taken as 1, as it converges to that value for
absorbing particles of radius $r \gtrsim$1 $\mu$m 
\citep[see e.g.][their Figure 5]{Moreno12a}. 

To make the problem tractable, a number of simplifying assumptions 
on the dust physical parameters must be made. Thus, the
particle density is taken as 1000 kg 
m$^{-3}$, and the geometric albedo is set to $p_v$=0.04, indicative of
dark material of carbonaceous composition  \citep[see
  e.g.][]{Moreno12a}. For the 
particle phase function correction, we use a linear phase coefficient 
of 0.03 mag deg$^{-1}$, which is in the range of comet 
dust particles in the 
1$^\circ \le \alpha \le$ 30$^\circ$ phase angle domain 
\citep[e.g.,][]{MeechJewitt87}. A
broad size distribution is assumed, with minimum
and maximum particle radii set to 10 $\mu$m and 1 cm, respectively, and 
following a power-law function of index $\kappa$=--3.2, which is in
the range  of previous estimates of the 
size distribution of particles ejected from activated asteroids and
comets. 

As the actual function describing the time evolution of the 
dust mass loss rate is highly uncertain, we simply assume that this is
given by a Gaussian function, for each 
component, with peak 
loss rate and time of maximum emission rate given by  $\dot{M}_0$, and $t_0$,
respectively. The full-width at 
half-maximum of the Gaussian, denoted as FWHM, gives a measure of the
effective time span of the emission event.  This parametrization
provides a fitting function with only three free parameters,
that has been otherwise proved useful to characterize the behavior of
other activated asteroids in the main belt \citep{Moreno16b,Moreno16c}.

The particles are assumed to be ejected isotropically from the
asteroid nuclei. We adopted a customary particle-size-dependent
velocity law parametrized as  
$v=v_0\beta^{1/\gamma}$, where  $v_0$ and $\gamma$ are fitting
parameters of the model. Since the ejection mechanism should be in
principle the
same for both fragments, we set the parameter $\gamma$ to be the same
for both asteroid components.   

In the modeling procedure, we have a total of nine fitting parameters: the 
three parameters associated to the dust loss rate function ($\dot{M}_0$,
$t_0$, and HWHM), one for each component, and the 
dust ejection velocity parameters $v_0$, one for each component,  
and $\gamma$, this
parameter being the same for both components. The model analysis, aimed
at finding the best-fit 
set of parameters, is conducted by the downhill
simplex method \citep{Nelder65}, using the FORTRAN implementation
described in \cite{Press92}. The quality of the fits is characterized
by minimizing the mean relative error of each model image as 
$\sigma_i=
\frac{\sum{\frac{|\log(I_{obs}(i))-\log(I_{mod}(i))|}{|\log(I_{obs}(i))|}}}{N(i)}$,
 where
$I_{obs}(i)$  and $I_{mod}(i)$ are the observed and modeled tail
brightness, and $N(i)$ is the number of pixels of image $i$. 
For the images in which the isophote field could not be
retrieved, this parameter is calculated as  
$\sigma_i=\sum\frac{|m_{obs}(i)-m_{mod}(i)|}{m_{obs}(i)}$, where
$m_{obs}(i)$  and $m_{mod}(i)$ are the measured and modeled magnitudes. The
fitting parameter is $\chi=\sum \sigma_i$, where the summation is
extended to all the images under consideration, i.e., $i$=1,8.

\section{Results and discussion}

The fitting of the images was accomplished by defining a set of five 
parameters per asteroid component, as stated in the previous
section. The best fit 
parameters, after running the code for a variety of different starting
simplexes, are shown in table 2. The derived 
synthetic R$_c$ magnitudes for each image are given in table 1, together
with the measured values. The resulting modeled isophotes are
displayed in figure 3.  
The agreement between the 
observations and the model isophotes and between
the measured and synthetic magnitudes is very good, the mean of the absolute
differences being only 0.07 mag. The uncertainties in the
determination of the best-fit 
parameters are calculated assuming a criterion for which a fit is 
not acceptable when $\chi$ exceeds 10\% of its best-fit value ($\chi$=0.069).

From the results obtained, we see that both fragments became active before
perihelion. The activation times are very  
similar, $\sim$--250 days to perihelion. The activity peak occurs very 
close to perihelion for the J1-A component, but nearly two months
before for the J1-B component. In both cases, the activity lasted 
several months, a typical behavior of main-belt comets. The integrated
ejected dust masses until the last observation of July 31, 2016, are
similar, with values of 
(8$\pm$2)$\times$10$^{6}$ kg, and (6$\pm$2)$\times$10$^{6}$ kg for 
J1-A and J1-B, respectively.

Combining the orbit dynamics results with the modeled activity, the
most likely scenario is that of a fragmentation event
during the previous asteroid orbit, whose fragments have become activated when
near perihelion in the current orbit.  The simultaneous activation
times for both components and the duration of the
activity, of at least 6-9 months, implies almost unambiguously that
ice sublimation is the responsible mechanism for the dust emission.  
Archival image search for the asteroid appearance during
the previous orbit, in particular during the 
perihelion passage, would be needed to confirm the fragmentation event.

\section{Conclusions}  

From the observations of this double-component, outer main-belt, asteroid P/2016
J1-A and J1-B, its orbital dynamics, and the dust tail modeling, 
the following conclusions can be drawn:  

1) The orbital dynamics computations on a large number (4$\times$10$^{4}$) of clone
asteroids, randomly chosen from the respective 
six-dimensional uncertainty ellipsoid in the orbital element space
around each of the nominal orbits of P/2016 J1-A
and J1-B components, reveal a fragmentation event that most likely
occurred $\sim$2300 days  
before the current orbit's perihelion. Thus, P/2016 J1 is a very 
remarkable case of an asteroid pair, being the youngest discovered so far.

2) From dust tail modeling, we conclude that both components become
active at nearly the same time $\sim$250 days before perihelion
passage. Both components display different evolution, with peak
emission rates at different times (near perihelion and $\sim$50 days to
perihelion for J1-A and J1-B), and total dust ejected of
(8$\pm$2)$\times$10$^{6}$ kg, and (6$\pm$2)$\times$10$^{6}$ kg,
respectively, until the latest observation of July 31, 2016.

3) The dust velocity parameters inferred are very similar for both
asteroidal components, with terminal velocities weakly dependent on the
particle size, and of order 0.6-0.9 m$^{-1}$ for the largest
particles ejected in the model. This is compatible with
the escape velocities expected from the maximum 
$\sim$500 m radius bodies in the 
1000 to 3000 kg m$^{-3}$ bulk density range inferred from their
absolute magnitudes, assuming a geometric albedo of 0.04.  

4) The most probable time of the closest approach between components and the start of the current
dust activity are separated approximately by one orbital period. 
Then, the most plausible scenario is that of a fragmentation of 
the parent asteroid in the previous orbit, whose
fragments have become activated nearly simultaneously in the present
perihelion approach. This, together with the long-standing activity
(6-9 months or longer) strongly suggests ice sublimation as the
responsible mechanism of the dust emission.

\acknowledgments

We are grateful to an anonymous referee for his/her 
comments and suggestions that helped to improve the
paper considerably.

This article is based on observations made with the Gran Telescopio
Canarias, installed in the Spanish Observatorio del Roque de los
Muchachos of the Instituto de Astrof\'\i sica de Canarias, in the island 
of La Palma, and on observations obtained with MegaPrime/MegaCam, a
joint project of CFHT and CEA/IRFU, at the Canada-France-Hawaii
Telescope (CFHT) which is operated by the National Research Council
(NRC) of Canada, the Institut National des Science de l'Univers of the
Centre National de la Recherche Scientifique (CNRS) of France, and the
University of Hawaii. This work is based in part on data products
produced at Terapix available at the Canadian Astronomy Data Centre as
part of the Canada-France-Hawaii Telescope Legacy Survey, a
collaborative project of NRC and CNRS.  

This work was supported by contracts AYA2015-67152-R and
AYA2015-71975-REDT from the Spanish
Ministerio de Econom\'\i a y Competitividad (MINECO). J. Licandro gratefully
acknowledges support from contract AYA2015-67772-R (MINECO, Spain). 

The Pan-STARRS1 Surveys (PS1) have been made possible through
contributions of the Institute 
for Astronomy, the University of Hawaii, the Pan-STARRS Project
Office, the Max-Planck Society and its participating 
institutes, the Max Planck Institute for Astronomy, Heidelberg and the
Max Planck Institute for Extraterrestrial Physics, 
Garching, The Johns Hopkins University, Durham University, the
University of Edinburgh, Queen's University Belfast, 
the Harvard-Smithsonian Center for Astrophysics, the Las Cumbres
Observatory Global Telescope Network Incorporated, 
the National Central University of Taiwan, the Space Telescope Science
Institute, the National Aeronautics and Space 
Administration under Grant Nos. NNX08AR22G, NNX12AR65G, and NNX14AM74G
issued through the Planetary Science Division 
of the NASA Science Mission Directorate, the National Science
Foundation under Grant No. AST-1238877, the University 
of Maryland, and Eotvos Lorand University (ELTE) and the Los Alamos National Laboratory.

B. Novakovi\'c also acknowledges support by the Ministry of
Education, Science and Technological Development of the Republic of
Serbia, Project 176011. 

B.T. Bolin is supported by l'\`{E}cole Doctorale Sciences
Fondamentales et Appliqu\'{e}es, ED.SFA (ED 364) at l'Universit\'{e}
de C\^ote d'Azur.

D. Farnocchia conducted this research at the Jet Propulsion
Laboratory, California Institute of Technology, under a contract with
NASA.

\clearpage

\begin{figure}[ht]
\centerline{\includegraphics[scale=0.8,angle=-90]{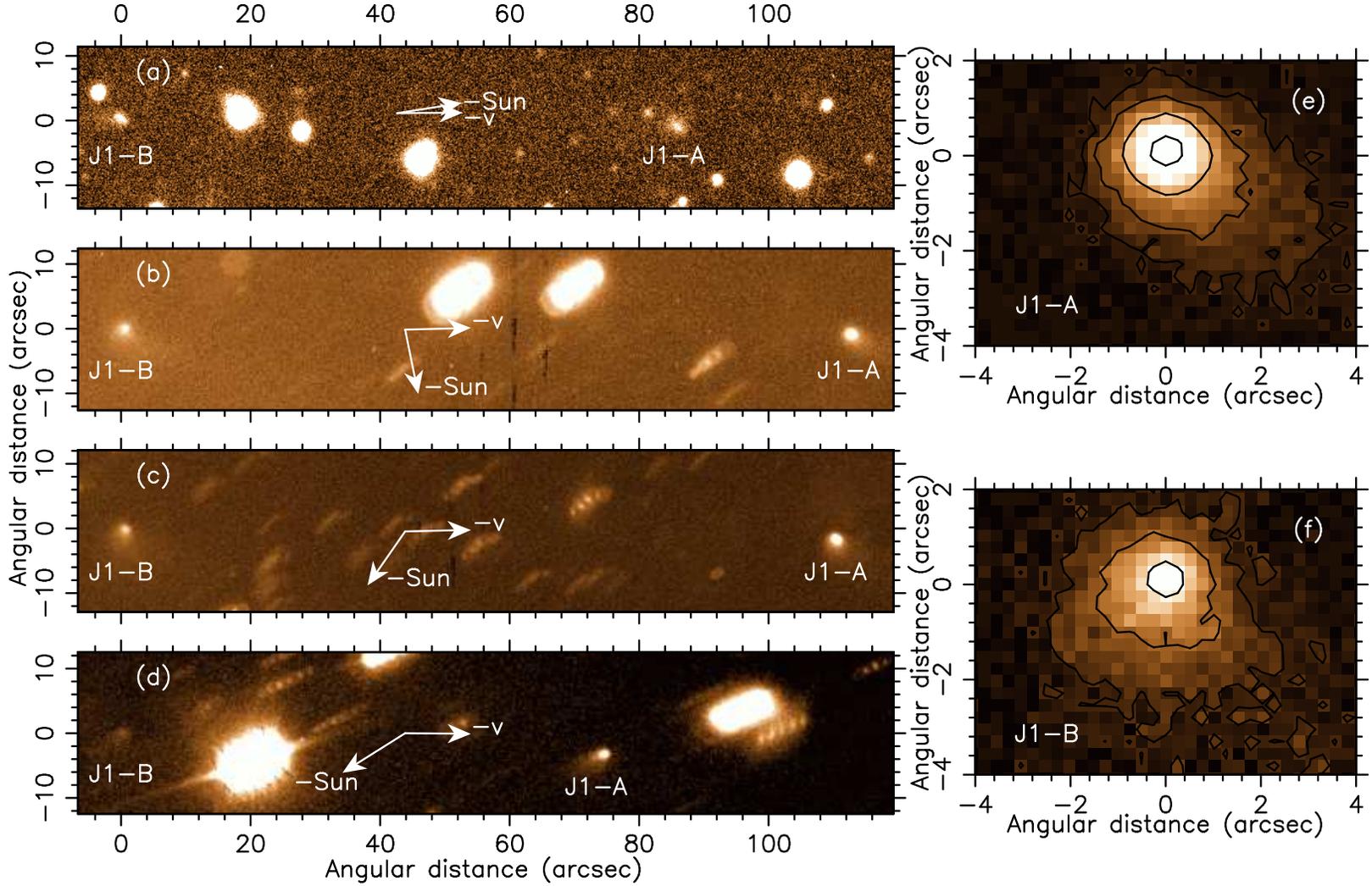}}
\caption{Images of P/2016 J1-A and J1-B obtained with MegaCam on the
  3.6m Canada-France-Hawaii-Telescope 
  on March 17, 2016 (a), and with OSIRIS at the 10.4m Gran Telescopio
  Canarias on May 15, 2016, May 29,
  2016, and July 31, 2016 (b,c,d). In panels (a),(b),(c), and (d), the physical
  dimensions are 170982$\times$33925, 133788$\times$26545,
  135616$\times$26908, and 184964$\times$36699 km, respectively.
Close-up views of J1-A and J1-B on May 15, 
  2016 are shown on panels (e), and (f),
  respectively. The innermost isophotes in panels (e) and (f) correspond to
  22.5 and 23.5 mag arcsec$^{-2}$ in the $r^\prime$ band,
  respectively. Isophotes increase in 
  steps of one magnitude outwards. North is up, and East is to the left in all
  panels. In panels (a) to (d), the projected directions opposite to
  the Sun and the negative of the 
  orbital velocity vectors are shown.
   \label{fig1}}
\end{figure}

\clearpage
\begin{figure}[ht]
\centerline{\includegraphics[scale=0.6,angle=-90]{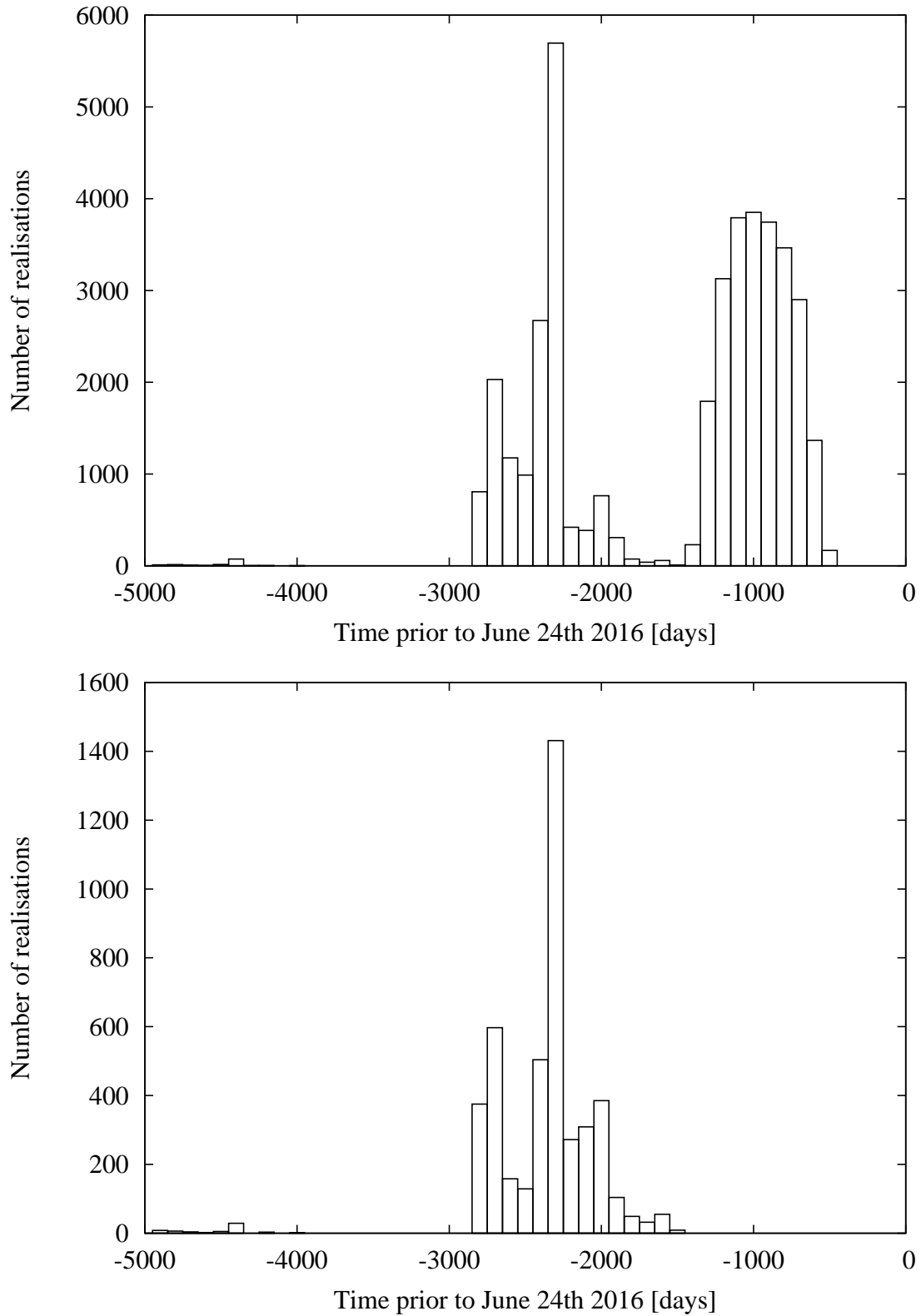}}
\caption{Upper panel: Frequency distribution of dates of closest
  approach between
  J1-A and J1-B clone pairs. Lower panel: same as in the upper panel, 
but only the pairs of clones that approached closer than 1500 km are shown.
    \label{figure2.eps}}
\end{figure}

\clearpage

\begin{figure}[ht]
\centerline{\includegraphics[scale=0.7,angle=-90]{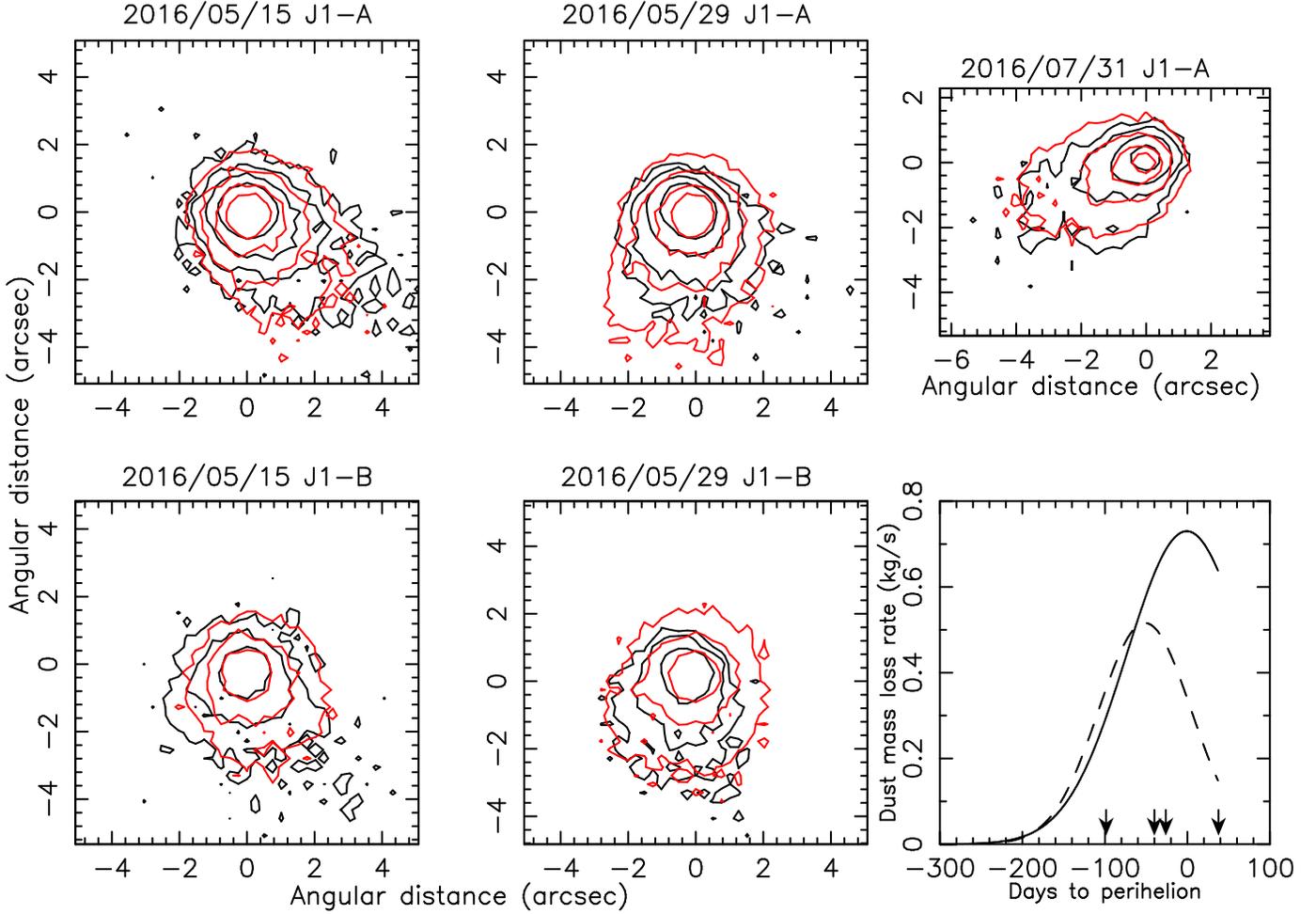}}
\caption{Measured isophotes (black contours) and best-fit model isophotes 
(red contours) for different dates, for J1-A and J1-B asteroid
  components. Innermost isophote levels are  2$\times$10$^{-14}$
  (J1-A, May 15 and May 29), 1.44$\times$10$^{-14}$ (J1-A, July 31), 
 1.25$\times$10$^{-14}$ (J1-B, May 15), and 10$^{-14}$ (J1-B, May
 29), all in solar disk intensity units. 
Isophotes decrease in factors of two outwards. Component J1-B on
 July 31 is not displayed, as it was too faint to build properly an
 isophote field. The lowermost right panel displays the best-fit dust
 loss rate as a function of time to perihelion for component J1-A
 (solid line) and J1-B (dashed line). Arrows indicate the observation
 dates (see table 1).}
\label{fig3}
\end{figure}

\clearpage

\begin{deluxetable}{ccccccccccc}
\rotate
\tablewidth{0pt}
\tablecaption{Log of the observations}
\tablehead{
\colhead{Observation date (UT)} & \colhead{Days to} &  \colhead{Total} & \colhead{$R_c$-mag}
&\colhead{$R_c$-mag} & \colhead{$R_c$-mag} & \colhead{$R_c$-mag} &
\colhead{R$^{1}$} & \colhead{$\Delta^{2}$} &  \colhead{$\alpha^{3}$}& \colhead{True} \\
\colhead{YYYY/MM/DD HH:MM} & \colhead{perihelion} &  \colhead{exposure} & \colhead{J1-A} & \colhead{J1-A} & \colhead{J1-B}
& \colhead{J1-B} 
& \colhead{(AU)} & \colhead{(AU)} &  \colhead{($^\circ$)}& \colhead{Anomaly}  \\
 & & time (s) & (measured) & (model) & (measured) & (model) & & & & ($^\circ$) \\}
\startdata
2016/03/17 14:53 &--98.6  & 300 & 23.23$\pm$0.12 & 23.23 &
23.02$\pm$0.14 & 23.02 & 2.501 & 1.871 & 20.5 & 332.3  \\
2016/05/15 02:11 &--40.1  & 900 & 20.53$\pm$0.03 & 20.81 &
20.80$\pm$0.04 & 20.88 & 2.457 & 1.464 & 5.65 & 348.6   \\
2016/05/29 01:02 &--26.2  & 900 & 20.66$\pm$0.03 & 20.73 &
20.99$\pm$0.05 &20.97 & 2.452 & 1.484 & 9.01 & 352.5  \\
2016/07/31 22:38 & +37.7  & 900 & 21.78$\pm$0.04 & 21.85 &
23.59$\pm$0.03 & 23.59 & 2.456 & 2.024 & 23.77 & 10.7  \\
\enddata
\tablenotetext{1}{Heliocentric distance}
\tablenotetext{2}{Geocentric distance}
\tablenotetext{3}{Solar phase angle}
\end{deluxetable}

\clearpage

\begin{deluxetable}{c|ccccccc}
\tablewidth{0pt}
\tablecaption{Best-fit parameters of the model for the two asteroid components}
\tablehead{
\colhead{P/2016 J1} & \colhead{$\dot{M}_0$}  & \colhead{t$_0$} & 
\colhead{FWHM} &  \colhead{v$_0$} &  \colhead{$\gamma$} &
\colhead{Total dust} \\ 

\colhead{component} & \colhead{(kg/s)} &\colhead{(days)} & 
\colhead{(days)} &  \colhead{(cm s$^{-1}$)} &
\colhead{ } & \colhead{mass ejected (kg)}  \\
}
\startdata
J1-A & 0.73$^{+0.15}_{-0.10}$ & -0.8$^{+6}_{-5}$  & 172$^{+11}_{-8}$
& 256$^{+100}_{-30}$ & 6.7$^{+0.5}_{-0.5}$ &  (8$\pm$2)$\times$10$^6$  \\
J1-B & 0.52$^{+0.10}_{-0.10}$ & --51.5$^{+4}_{-6}$  & 132$^{+8}_{-7}$
& 343$^{+40}_{-30}$ & 6.7$^{+0.5}_{-0.5}$ &  (6$\pm$2)$\times$10$^6$  \\
\enddata
\end{deluxetable}

\end{document}